\documentclass[epjCONF,columns]{svjour} 
\usepackage{graphicx}
\usepackage{multirow}
\usepackage{subfigure}
\usepackage[varg]{txfonts} 
\usepackage[applemac]{inputenc}
\usepackage{lineno}
\def\coupling{\ensuremath{k/\overline{M}_{Pl}}}
\session-title{Hadron Collider Physics Symposium 2011}
\begin{document}
%
\title{Search for extra dimensions in the diphoton final state with ATLAS}
\author{Quentin Buat\thanks{\email{quentin.buat@cern.ch}} On behalf of the ATLAS Collaboration}
\institute{LPSC-Grenoble, CNRS/IN2P3, UJF, INPG}
\abstract{
The large difference between the Planck scale and the electroweak scale, 
known as the hierarchy problem, has been addressed in some models 
through the existence of extra spatial dimensions. 
A search for evidence of extra spatial dimensions has been performed, 
through an analysis of the diphoton final state in data recorded in 2011 
with the ATLAS detector at the CERN Large Hadron Collider. 
The analysis uses a dataset of 2.12 $fb^{-1}$ 
of proton-proton collisions at $\sqrt{s} = 7$ TeV. 
The diphoton invariant mass spectrum is observed to be 
in good agreement with the expected Standard Model (SM) background. 
We set 95$\%$ CL lower limits on the scale related to virtual graviton exchange process 
in the context of the Arkani-Hamed, Dimopoulos, Dvali model (ADD) 
and on the lightest Kaluza Klein excitation mass in the context of 
the Randall-Sundrum model (RS). 
} 
\maketitle
\section{Introduction}
\label{sec:intro}
Recently, there has been great interest in models which 
address the hierarchy problem through the existence of extra spatial dimensions. In this
analysis we search for evidence of extra spatial dimensions in the context of the ADD 
\cite{ADDPaper} and RS \cite{RSPaper} models. This analysis briefly summarized here, 
is described in more detail in \cite{GravPaper}.

In the ADD context, \emph{n} flat extra spatial dimensions are postulated with a compactification radius 
R. The gravity is the only field that propagates in those extra dimensions and acquires Kaluza Klein (KK) modes. 
In the ADD context, resolving the hierarchy problem implies small values of 1/R. This leads to an 
almost continuous KK spectrum of the graviton mass. Experimentally, an ADD signal will contribute through virtual
graviton exchange to the diphoton invariant mass spectrum. This process can be parametrized as a function of the number
of extra dimensions \emph{n} and an ultraviolet cutoff ($M_s$). Several formalisms
exist in the literature to define $M_s$, referred to here as GRW \cite{GRW}, 
HLZ \cite{HLZ} and Hewett \cite{Hewett}.

The RS model postulates a 5-dimensional space-time bounded by two (3+1) branes with the SM 
particles localized on one of the branes. The fifth dimension has a "warped" geometry which allows 
to naturally generate TeV scales from the Planck scale. The fundamental Planck scale  ($M_{Pl}$) 
on one brane is related to the apparent scale ($\Lambda_{\pi}$) on the other brane by the relation  
$\Lambda_{\pi} =\overline{M}_{Pl} \exp{(-k\pi r_c)}$ where $k$ is the curvature scale of the extra  dimension, 
$r_c$ the compactification radius and  $\overline{M}_{Pl}=\frac{M_{Pl}}{\sqrt{8\pi}}$. The observed hierarchy 
of scales can be naturally reproduced by this model with $kr_c\approx 11$. In the minimal RS model, 
gravitons are the only particles that propagate in the bulk. Consequently a series of massive KK excitations 
is predicted with a mass splitting of the order of 1 TeV. Finally the RS model can be expressed in terms of the 
coupling $\coupling$ and the mass of the lightest KK excitation ($m_G$). Experimentally the presence of a RS
 signal can be probed as a narrow resonance in the diphoton invariant mass spectrum.
\section{Photon reconstruction and identification}
\label{sec:AtlasDetector}
A complete description of the ATLAS detector can be found in \cite{DetectorPaper}. In this analysis 
we look for final states with two photons. Therefore we rely on the inner tracker and the calorimetric 
system of the detector. The tracking system of \linebreak[4] ATLAS is composed of layers of silicon-based and 
straw-tube detectors surrounded by a 2 T magnetic field for momentum measurements. It allows to 
measure the tracks of the 30\% of photons converting in $e^+e^-$ pairs before reaching the calorimeter.
The photon energy deposit is measured by the Liquid Argon (LAr) electromagnetic calorimeter, 
covering the region $0<|\eta|<1.37$ and $1.52<|\eta|<2.37$. It is segmented in three 
longitudinal layers. Most of the energy of the electromagnetic (EM) shower is recovered by the second 
layer which has a granularity of $\Delta\eta\times\Delta\phi = 0.025\times0.025$. The first layer with a thinner 
segmentation is designed to reject jets dominated by neutral hadrons such as a $\pi^0$ or a $\eta$. 
Indeed, such a hadron may decay into two photons which can be identified by resolving the two maxima 
in the first layer of the LAr calorimeter. In the central region ($|\eta|<1.81$), a presampler is used to 
evaluate the energy lost in the upstream material. The hadronic calorimeter, composed of scintillating 
tiles-iron sampling for the central part ($|\eta|<1.7$) and of liquid-argon-copper/tungsten sampling for 
$|\eta|>1.7$ is used to measure the leakage of energy of the photon candidates. A complete description 
of the reconstruction and identification of photons with \linebreak[4] ATLAS can be found in \cite{ATLASphotons}. 
\section{Trigger and Data selection}
\label{sec:TriggerDataSel}
The events are recorded using a trigger requiring at least two photon candidates with transverse
 energy $E_T^\gamma > 20$ GeV and satisfying  identification requirements based on the leakage 
 of energy into the hadronic calorimeter and the transverse width of the EM shower. 
This trigger is fully efficient for high mass diphoton events passing the offline selection requirements.
Events are required to have at least one primary collision vertex, with at least three reconstructed tracks. 
Selected events have at least two photon candidates, each with $E_T^{\gamma} > 25$ GeV and lying outside
the transition region between the barrel and endcap calorimeters. In addition, the two photons are required to satisfy 
standard quality criteria and to lie outside detector regions where their energy is not measured in an optimal way.
The two highest $E_T^{\gamma}$ photon candidates have to satisfy a set of identification requirements on the 
hadronic leakage and on the lateral width of the EM shower. The requirements on the EM shower use the thin 
granularity of the first sampling to achieve a high purity of the selected photon sample. The isolation transverse 
energy $E_T^{iso}$ for each photon is calculated~\cite{ATLASphotons} by summing over the cells within a cone of
radius $\Delta R = \sqrt{(\eta - \eta^\gamma)^2 + (\phi - \phi^\gamma)^2} < 0.4$ around the direction of the photon. 
Then the energy deposit of the photon itself is subtracted as well as the soft jet activity of the underlying event 
\cite{ambientenergy}. In addition, $E_T^{iso}$ is corrected for the leakage of the photon energy into the isolation ring. 
To further reduce the jet background, an isolation cut of $E_T^{iso} < 5$ GeV  is applied on the two leading photons.
For all events the two photons with the highest $E_T^\gamma$ values are considered and  the diphoton invariant 
mass has to exceed 140 GeV. A total of 6846 events are selected.
\section{Monte Carlo Simulation Studies}
\label{sec:Sim}
Monte Carlo (MC) simulations were performed to study the detector response to various scenarios of 
ADD and RS models as well as to perform studies of the SM background. 

The SM diphoton background was simulated with \linebreak[4] PYTHIA 6.424 \cite{pythia} and MRST2007LOMOD 
\cite{MRST2007lomod} parton distribution functions (PDFs). The PYTHIA events were 
reweighted as a function of the diphoton invariant mass to the differential cross 
section predicted by the NLO calculation of DIPHOX 1.3.2 \cite{diphox} . The reweighting 
factor decreased smoothly from $\approx 1.6$ for $m_{\gamma\gamma} = 140$ GeV to 1
 for large masses. 
 
 SHERPA 1.2.3 \cite{sherpa} with CTEQ6L~\cite{cteq66} PDFs was used 
 to simulate the various ADD scenarios. The ADD MC samples were used to determine the
signal acceptance ($A$) and selection efficiency ($\epsilon$).  The acceptance varied for the
various ADD implementations and fell from typical values of $\approx 20$\% for $M_S = 1.5$ TeV
down to $\approx 15$\% for $M_S = 3$ TeV. The selection efficiency, for events within the detector 
acceptance, was found to be $\approx 70$\%. 

RS model MC signal samples were produced with \linebreak[4] PYTHIA 6.424 
for a range of $m_G$ and \coupling \ values, using the MRST2007LOMOD PDFs. 
The products of \linebreak[4] $A\times\epsilon$ for the RS signal models 
were in the range $\approx (53-60)$\%, slowly rising with increasing graviton mass.
The reconstructed shape of the graviton resonance was modeled by a Breit-Wigner (BW) lineshape convoluted
with a double-sided Crystal Ball function to describe the detector response.
The natural width of the BW was fixed according to the expected theoretical value, which
varies as the square of \coupling.  As an example, the value of the width for $\coupling = 0.1$
increases from $\approx 8$ GeV for $m_G=800$ GeV to $\approx 30$ GeV for $m_G=2200$ GeV. 
Fitting this mass dependence provides a parametrization used to describe signals with any values 
of $m_G$ and \coupling.
\section{Background evalution}
\label{sec:Bkg}
The SM $\gamma\gamma$ production is the largest background of this analysis. 
The invariant mass spectrum shape was determined by PYTHIA reweighted by DIPHOX 
NLO cross-section predictions. 

The second significant background arises from a different 
physics object (electron or jet) misidentified as a photon. This background, called reducible 
background, is dominated by the production of $\gamma$ + jet or dijet events. Studies have 
shown that the Drell Yan contribution can be neglected for events with 
$m_{\gamma\gamma} > 140$ GeV. The shape of the reducible background is determined 
by a data-driven technique. By reverting the identification criteria on the leading and/or subleading 
photon, we get three different control samples  enriched in $\gamma$ + jet, jet + $\gamma$ and 
dijet events. The invariant mass spectra of the three samples are consistent and they 
are merged to determine the reducible background shape. This shape is extrapolated to high 
diphoton masses where the data control sample are statistically limited. 

The fraction of each background has been determined by a two dimensional template fit method using 
the photon isolation variable, described in more detail in \cite{SMdiphoton}. Finally the background 
expectation is normalized to the data in the invariant mass range [140, 400] GeV where the presence 
of any possible ADD and RS signal has been excluded by previous searches.
\section{Systematic uncertainties}
\label{sec:SysUnc}
The systematic uncertainty on the background estimation arises from three different sources. 
The irreducible background uncertainty is obtained by varying the scales of the models and the 
PDFs in DIPHOX while the reducible background uncertainty is obtained by fitting the three control samples 
described in section \ref{sec:Bkg} with the functionnal form. The uncertainty on the background estimation 
varies from $\approx$2\% for the low mass region to $\approx$20\% at a mass of $\approx$2 TeV. 
For the signal, the PDFs uncertainty is of 10-15\% for ADD models and 5-10 \% for RS models and the 
uncertainty on the signal yields has been evaluated to be of 6.7 \% for both ADD and RS models.
\section{Results and Interpretation}
\label{sec:Res}
\begin{figure}[h!]
\resizebox{\columnwidth}{!}{
\includegraphics{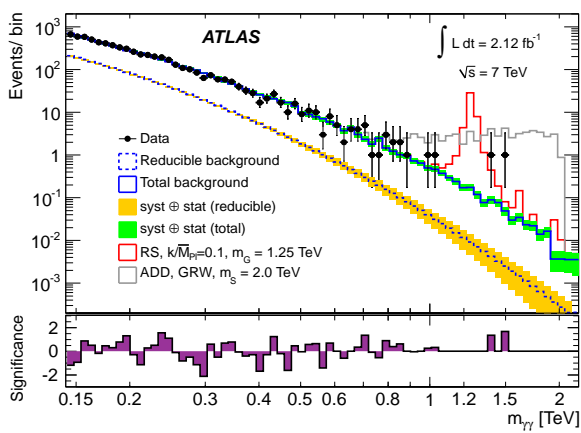} }
\caption{Observed invariant mass spectrum, superimposed with the predicted 
SM background and examples of signals for ADD and 
RS models. 
The bin-by-bin significance of the difference 
between data and background is shown 
in the lower panel.}
\label{fig:mgg}      
\end{figure}

\begin{figure}[h!]
\centering
\resizebox{0.75\columnwidth}{!}{
\subfigure{ \includegraphics{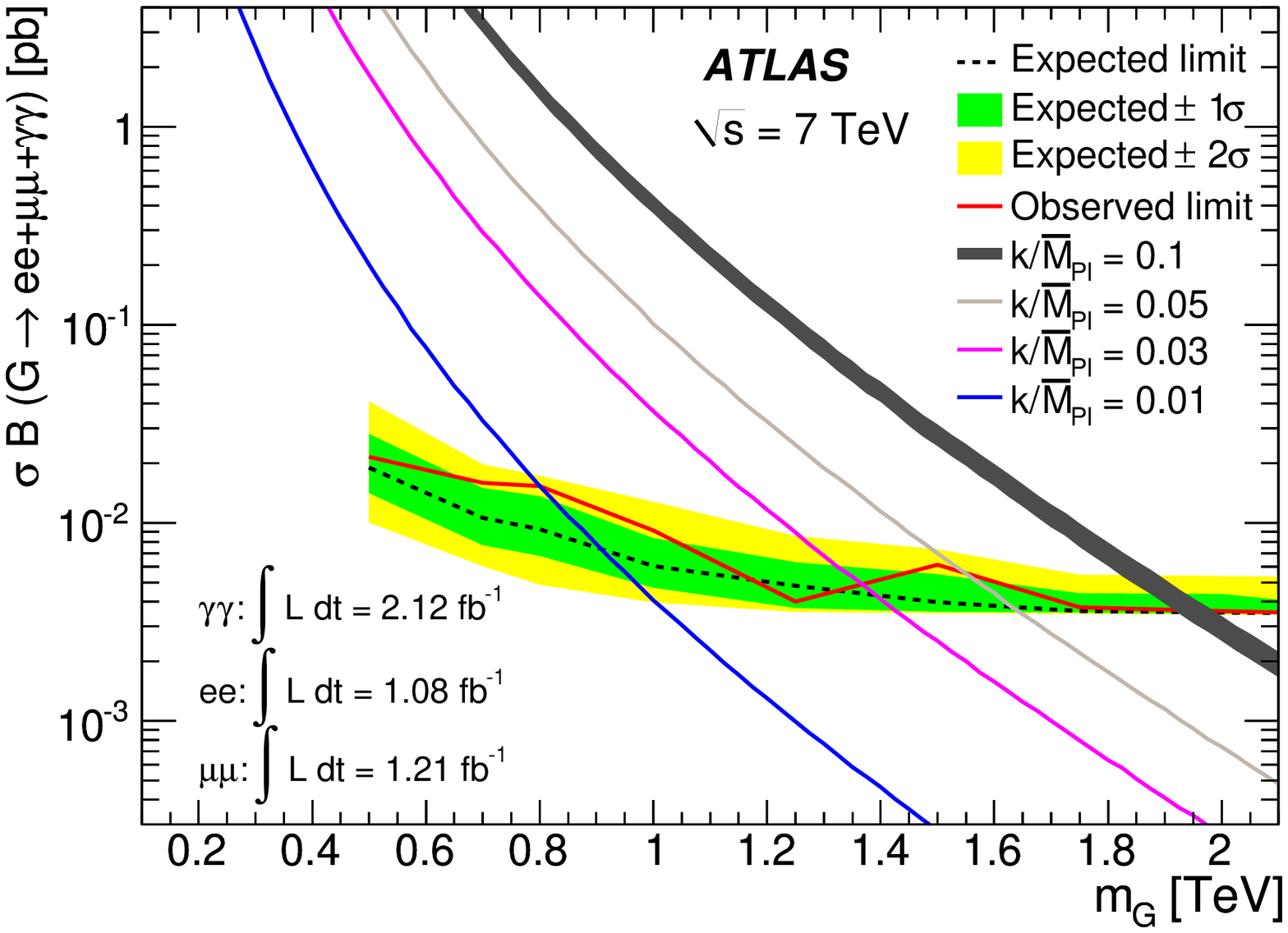} }
}
\resizebox{0.75\columnwidth}{!}{
\subfigure{\includegraphics{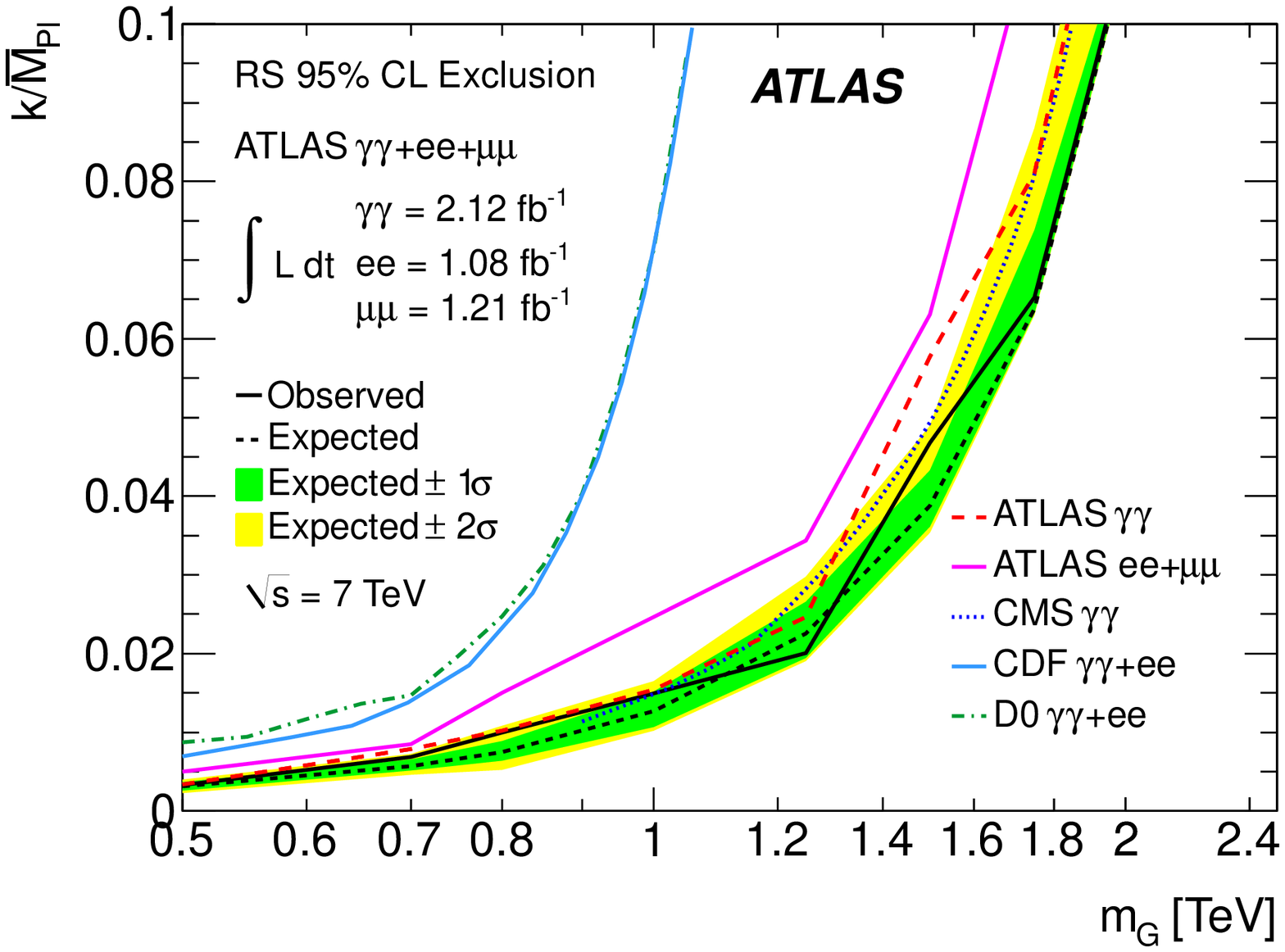} }
}
\caption{(Top) Expected and observed 95\% CL limits 
from the combination of $G \rightarrow \gamma\gamma/ee/\mu\mu$
channels on $\sigma B$ as a function of the graviton mass.
The theory bands are drawn assuming a k-factor of 1.75. 
 (Bottom) The RS results interpreted in 
the plane of \coupling \ versus graviton mass,
and including recent results from 
other experiments~\cite{TevatronRS,CMSggnew}.
The region above the curve is excluded at 95\% CL. 
}
\label{fig:limits}      
\end{figure}

Figure \ref{fig:mgg} shows the observed invariant mass spectrum of diphoton events, with 
statistical significance of the bin-by-bin difference between data and the expected background 
at the bottom. This difference is measured in standard deviations based on Poisson statistics. 
The small variations show the good agreement between data and the estimated SM background. 
This is confirmed by an analysis using the BUMPHUNTER \cite{bumphunter}  tool, which yields a probability of 
0.28, given the background only hypothesis, to observe discrepancies as 
large as the one observed in the bin by bin comparison. In the absence of any significant deviation 
we set 95\% CL  on the signal cross sections, using a bayesian approach \cite{bayes} with a flat prior.
In the context of ADD models the signal search region was chosen as $m_{\gamma\gamma}> 1.1$ TeV 
by optimizing the expected limit. The observed (expected) limit on the cross section due to new physics 
is 2.49 (1.94) fb. This result can be translated into 95\% CL lower limits on $M_s$ for different numbers 
of extra dimensions and formalisms. Table \ref{tab:ADDlimits} summarizes the observed limits value.
In the context of the RS models, the observed invariant mass spectrum was compared to templates of 
the expected background and various signal parametrizations. The limit is set as 95\% CL on the product 
of the cross-section $\times$ Branching ratio  ($\sigma B$) as a function of $m_G$. Then the cross-section 
limit is converted into a mass limit using the theoretical dependence. Table \ref{tab:RSlimits} shows 
the limit in the diphoton final state for various values of \coupling. By combining with previously published 
ATLAS results \cite{zprime} from the dilepton final state, we obtain a 95\% CL limit on $\sigma B$ as a 
function of $m_G$ shown by the top plot of figure \ref{fig:limits}. The result is also interpreted in the 
plane of \coupling \ versus $m_G$ (bottom plot of figure \ref{fig:limits}).

\begin{table}\scriptsize
\caption{95\% CL limits on $M_S$ (in TeV) 
for  various implementations of the ADD model,
using both LO (k-factor = 1) and NLO (k-factor = 1.70) 
theory cross section calculations.}
\label{tab:ADDlimits}       
\addtolength{\tabcolsep}{-1pt}
\begin{tabular}{|c|c|cc|ccccc|} 
\hline
 k-factor & GRW & \multicolumn{2}{c|}{Hewett} & \multicolumn{5}{c|}{HLZ} \\ 
  Value    &     & Pos & Neg &$n=3$&$n=4$&$n=5$&$n=6$&$n = 7$\\ \hline \hline
 1  & 2.73 & 2.44 &  2.16 & 3.25 & 2.73 & 2.47 & 2.30 & 2.17 \\ \hline
1.70 &  2.97 & 2.66 & 2.27 & 3.53 & 2.97 & 2.69 & 2.50 & 2.36  \\ \hline 
\end{tabular}
\end{table}

\begin{table}\scriptsize
\caption{95\% CL lower limits on $m_G$, for  various
values of \coupling. The  $G\rightarrow\gamma\gamma$ channel alone and 
the combination with the dilepton results of Ref.~\cite{zprime} are shown, 
using both LO (k-factor = 1) and
NLO (k-factor = 1.75) theory cross section 
calculations.}
\label{tab:RSlimits}       
    \begin{tabular}{|c|c| c c c c|} \hline
      \multirow{3}{45pt}{\centering{k-Factor Value}}  & \multirow{3}{60pt}{\centering{Channel(s) Used}} & \multicolumn{4}{c|}{95\% CL Limit [TeV]} \\ 
      \cline{3-6}
      &                                  & \multicolumn{4}{c|}{ $k/\overline{M}_{Pl}$ Value}  \\
      &                                             & 0.01     &0.03      &0.05        &0.1            \\ 
      \hline\hline
      \multirow{2}{*}{1}     &$G\to\gamma\gamma$    &0.74      &1.26      &1.41        &1.79  \\
      & $G\to \gamma\gamma$/$ee$/$\mu\mu$           &0.76      &1.32      &1.47        &1.90  \\ 
      \hline \hline
      \multirow{2}{*}{1.75} & $G\to\gamma\gamma$    &0.79      &1.30      &1.45        &1.85  \\
      &$G\to \gamma\gamma$/$ee$/$\mu\mu$            &0.80      &1.37      &1.55        &1.95  \\
      \hline
    \end{tabular}
\end{table}

\end{document}